\documentclass[manuscript]{aastex63}

\usepackage[whole]{bxcjkjatype}

\received{October 19, 2021}
\revised{January 28, 2022}
\accepted{February 14, 2022}

\shorttitle{Orbital Evolution of Close-in Super-Earths}
\shortauthors{Fujita et al.}

\begin{document}

\title{Orbital Evolution of Close-in Super-Earths Driven by Atmospheric Escape}

\correspondingauthor{Naho Fujita}
\email{fujita@kusastro.kyoto-u.ac.jp}

\author[0000-0002-5791-970X]{Naho Fujita}
\affiliation{Department of Astronomy, Kyoto University, Kitashirakawa-Oiwake-cho, Sakyo-ku, Kyoto 606-8502, Japan}

\author[0000-0003-4676-0251]{Yasunori Hori}
\affiliation{Astrobiology Center, 2-21-1 Osawa, Mitaka, Tokyo 181-8588, Japan}
\affiliation{National Astronomical Observatory of Japan, 2-21-1 Osawa, Mitaka, Tokyo 181-8588, Japan}

\author[0000-0003-1242-7290]{Takanori Sasaki}
\affiliation{Department of Astronomy, Kyoto University, Kitashirakawa-Oiwake-cho, Sakyo-ku, Kyoto 606-8502, Japan}

\begin{abstract}
The increasing number of super-Earths close to their host stars revealed a scarcity of close-in small planets with 1.5--2.0\,$R_\oplus$ in the radius distribution of {\it Kepler} planets. The atmospheric escape of super-Earths by photoevaporation can explain the origin of the observed ``radius gap.''
Many theoretical studies considered the in-situ mass loss of a close-in planet. Planets that undergo the atmospheric escape, however, move outward due to the change in the orbital angular momentum of their star-planet systems. In this study, we calculate the orbital evolution of an evaporating super-Earth with a H$_2$/He atmosphere around FGKM-type stars under a stellar X-ray and extreme UV irradiation (XUV).
The rate of increase in the orbital radius of an evaporating planet is approximately proportional to that of the atmospheric mass loss during a high stellar XUV phase. We show that super-Earths with a rocky core of $\lesssim 10\,M_\oplus$ and a H$_2$/He atmosphere at $\lesssim$ 0.03--0.1\,au ($\lesssim$ 0.01--0.03\,au) around G-type stars (M-type stars) are prone to the outward migration driven by photoevaporation.
Although the changes in the orbits of the planets would be small, they would rearrange the orbital configurations of compact, multi-planet systems, such as the TRAPPIST-1 system.
We also find that the radius gap and the so-called ``Neptune desert'' in the observed population of close-in planets around FGK-type stars still appear in our simulations. On the other hand, the observed planet population around M-type stars can be reproduced only by a high stellar XUV luminosity model.
\end{abstract}


\section{INTRODUCTION} \label{sec:intro}

Since the first discovery of a hot Jupiter orbiting a Sun-like star was reported in 1995 \citep{1995Natur.378..355M}, over 4500 exoplanets have been discovered. Exoplanet statistics show that super-Earths account for the majority of exoplanetary systems \citep[e.g.,][]{2011arXiv1109.2497M,2012ApJS..201...15H}, especially around low-mass stars \citep{2013A&A...549A.109B,2015ApJ...807...45D}.
Interestingly, there is a lack of close-in small planets with 1.5--2.0\,$R_\oplus$ and orbital periods of $< 100$\,days in the radius distribution of {\it Kepler} planets around FGK-type stars \citep{2017AJ....154..109F,2018MNRAS.479.4786V}, and M-type stars \citep{2021MNRAS.507.2154V}.
This ``radius gap/valley'' is understood as the radius boundary between bare rocky planets \citep{2014ApJ...783L...6W, 2015ApJ...801...41R} and planets with a substantial amount of atmosphere (at most $\sim$ 1--10\,\% of their total masses)  \citep[e.g.,][]{2015ApJ...806..183W}.
The transmission spectra of close-in super-Earths above the radius gap, such as GJ 1214b \citep[e.g.,][]{2014Natur.505...69K} and GJ 3470b \citep[e.g.,][]{2019NatAs...3..813B}, revealed that they have either a (hazy/cloudy) hydrogen-rich atmosphere or a metal-rich atmosphere.
The hydrogen-rich atmosphere of a super-Earth can be interpreted as the remnant of an accreted disk gas, namely, the primordial atmosphere.
A metal-rich atmosphere of a super-Earth is likely to originate from post-formation processes such as volcanic activities and outgassing from the impact material.

Super-Earths close to a host star are in danger of atmospheric loss under an intense stellar radiation field.
There are two main mechanisms that cause planets to lose their atmospheres\footnote{The silicate mantle and iron core inside a super-Earth can play a role in the hydrogen budget. The equilibrium chemistry between an interior and a hydrogen-rich atmosphere of a super-Earth can reduce the hydrogen mass fraction in the atmosphere \citep[e.g., see][]{2021arXiv210710405S}.}. This can occur through (1) ion pick-up by high-energy charged particles from a stellar wind or coronal mass ejection \citep[e.g.,][]{2015ApJ...813...50K}, and (2) thermal processes in the upper layer driven by X-ray or extreme-UV (XUV) heating, e.g., the Jeans escape and hydrodynamic escape \citep{1981Icar...48..150W}.
For a non-magnetized or weakly magnetized planet, like Mars, the non-thermal process strongly affects atmospheric mass loss.
Although a super-Earth may have a weak and short-lived magnetic field \citep[][]{2010ApJ...718..596G,2011ApJ...726...70T}, the lifetime of a dipole magnetic field can be as long as $\sim 1$\,Gyr for a rapidly rotating super-Earth \citep{2012Icar..217...88Z}. The magnetic protection against the invasion of high-energy charged particles into a close-in super-Earth is likely maintained for $\sim 1$\,Gyr \citep{2013ApJ...770...23Z}.
Thus, the XUV-driven atmospheric escape is considered the dominant mechanism for sculpting the atmospheres of close-in super-Earths.
In fact, the atmospheric escape of super-Earths by photoevaporation can explain the radius gap observed in the population of {\it Kepler} planets \citep[e.g.,][]{2017ApJ...847...29O}.

Atmospheric escape from a planet causes a change in its orbital radius.
A planet that loses its mass should move outward\footnote{Note that a close-in planet also experiences orbital decay by tidal torques \citep[e.g.,][]{2016CeMDA.126..227J,2017MNRAS.469..278G}.} owing to the change in the orbital angular momentum.
Pioneer studies investigated the orbital evolution of a planet that undergoes its atmospheric loss and gas drag due to a stellar wind from an evolved star, using simplified expressions \citep[e.g.,][]{1984MNRAS.208..763L}.
Subsequently, \citet{2007ApJ...661.1192V,2009ApJ...705L..81V} studied how a gas giant survives around an evolved star, including the orbital change of a planet driven by photoevaporative mass loss.
These studies considered the isotropic mass loss from a planet driven by stellar XUV irradiation. 
Photoevaporative mass loss from an irradiated planet, however, may be associated with an anisotropic wind driven by the day-night heating contrast \citep[e.g.][]{2009ApJ...694..205S,2013ApJ...766..102G} and nonspherically symmetric outflow controlled by magnetic fields \citep{2011ApJ...730...27A}.
Anisotropic ejection by evaporation of a hot Jupiter \citep{2012A&A...537L...3B} and a super-Earth/hot Neptune \citep{2015MNRAS.452.1743T} can induce orbital migration, depending on the geometry of the mass ejection, such as the ejection direction of the material and the orbital velocity.
The mass transfer from a planet to its host star via Roche lobe overflow also causes its orbital expansion.
\citet{1998ApJ...500..428T} investigated the effect of mass loss via the Roche lobe overflow on the orbital evolution of a migrating gas giant. \citet{2014ApJ...793L...3V} demonstrated the orbital expansion of a close-in gas giant accompanied by Roche lobe overflow in order to explain the observed population of small ultra-short-period planets \citep[][]{2014ApJ...787...47S, 2016PNAS..11312023S}.
Recently, \cite{2016CeMDA.126..227J} formulated the orbital evolution of a gas giant due to the mass loss via Roche lobe overflow, incorporating the effect that part of the orbital angular momentum of the atmosphere flowing out of a planet remains in the star-planet system.
Thus, the orbital evolution of a gas giant that undergoes mass loss has been discussed in the context of either stellar evolution or Roche lobe overflow.

Many theoretical studies on the photoevaporation of super-Earths, however, assume that they stay in situ while losing their atmospheres for $\sim$\,Gyr. Note that most of the evaporation driven by stellar XUV irradiation occurs in the first few 100\,Myr.
The strength of a stellar XUV flux that a planet receives determines its mass loss rate. As a planet migrates outward, a reduced stellar XUV irradiation should mitigate mass loss from the planet; that is, the atmospheric escape from a super-Earth and its orbital evolution occur simultaneously.
Therefore, the co-evolution of the orbit and the atmosphere of an evaporating, close-in planet is crucial to better understanding the formation history and orbital evolution of super-Earths under atmospheric conditions.

In this study, we calculate the orbital evolution of close-in super-Earths that undergo atmospheric loss driven by stellar XUV irradiation. Using a one-dimensional model for the mass loss from a super-Earth, we investigate how the angular momentum change due to atmospheric escape influences its orbital migration.
The remainder of this paper is structured as follows. In Section \ref{sec:methods}, we describe our model of the atmospheric escape and orbital evolution of a super-Earth. In Section \ref{sec:results}, we demonstrate that the atmospheric escape from a close-in super-Earth never strands its orbit and strongly affects the final orbital configuration of the planetary systems. We also compare the period-radius distribution of the simulated planets with that of the observed planets. In Section \ref{sec:discussions}, we discuss the impact of the geometry of an evaporative wind, tidal torques, and the eccentricity on the outward migration of close-in planet systems. Finally, in section \ref{sec:summary}, we summarize our results.

\section{METHODS} \label{sec:methods}

Close-in super-Earths experience atmospheric escape driven by stellar XUV irradiation. 
The change in the orbital angular momentum of a planet due to atmospheric mass loss causes outward migration.
We consider the orbital evolution of close-in super-Earths with a H$_2$/He atmosphere onto a rocky core of 0.5--20\,$M_\oplus$ around FGKM-type stars (0.2--1.4\,$M_\odot$) for 1\,Gyr after disk dispersal. The initial mass fraction of a H$_2$/He atmosphere relative to the core mass ranged from 1\,\% to 10\,\%.

\subsection{Stellar Evolution}

We considered a close-in planet with a H$_2$/He atmosphere around FGKM-type stars. 
The evolution of a planet-hosting star was simulated using the general-purpose stellar evolution code {\tt MESA} \citep{2011ApJS..192....3P} to derive the stellar intrinsic luminosity, temperature, and radius. The time evolution of the stellar XUV luminosity was estimated from the X-ray-to-bolometric luminosity relations of stars in open clusters aged $5-740\,\mathrm{Myr}$,
\begin{equation}
  \frac{L_\mathrm{XUV}}{L_\mathrm{bol}} = \left\{
  \begin{array}{ll}
    (L_\mathrm{XUV}/L_\mathrm{bol})_\mathrm{sat} & \; \mathrm{for} \;\; t \leq \tau_\mathrm{sat}\\
    (L_\mathrm{XUV}/L_\mathrm{bol})_\mathrm{sat} (t/\tau_\mathrm{sat})^{-\alpha} & \; \mathrm{for} \;\; t > \tau_\mathrm{sat},
  \end{array}
  \right.
\end{equation}
where $t$ is the age of the star ($t = 0$ is defined as the time when the star enters the pre-main sequence phase.), $\tau_\mathrm{sat}$ is the saturation turn-off time, $L_\mathrm{bol}$ is the stellar bolometric luminosity, and $\alpha$ is the power law index of $0.790 \leq (B-V)_0 < 0.935$, $0.565 \leq (B-V)_0 < 0.675$, and $0.290 \leq (B-V)_0 < 0.450$ for K-,G-, and F-type stars, respectively \citep{2012MNRAS.422.2024J}.
The $(B-V)$ color index of a star varies with time. In this study, we calculate the orbital evolution of close-in super-Earths via photoevaporation after disk dispersal ($t = 0.04$\,Gyr) until $t = 1$\,Gyr.
Since we do not consider the pre-main-sequence phase of the host star,
we adopt the $(B-V)$ value of a main-sequence star with a given stellar mass for the stellar XUV luminosity.
The upper limit values for $\log{(L_\mathrm{XUV}/L_\mathrm{bol})}$ given by \citet{2012MNRAS.422.2024J} were used in our simulations. The XUV luminosity of M-type stars follows that of \citet{2008A&A...479..579P}, where the XUV luminosity for $t < 0.6$\,Gyr is assumed to be at the saturation level.

\subsection{Atmospheric Escape}

We consider two regimes of atmospheric escape driven by stellar irradiation \citep{2009ApJ...693...23M,2012MNRAS.425.2931O}, an X-ray-driven radiation recombination-limited regime, and an EUV-driven energy-limited mass loss.
A high stellar XUV flux fully photoionizes the hydrogen atom in the upper layer. The temperature in the X-ray heated region is almost constant ($\sim 10000$\,K) because the radiative cooling due to H$_3^+$ ion and Lyman $\alpha$ emission from the recombination of hydrogen ions and electrons can be balanced with the photoionization of hydrogen. The mass flux of the X-ray heated flow (i.e., isothermal wind) passing through the sonic point is given by
\begin{equation}
    \dot{M}_{p} = -4\pi \rho_\mathrm{s} c_\mathrm{s} r^2_\mathrm{s},
      \label{eq:mass-loss-rc}
\end{equation}
where the subscript ``$\mathrm{s}$'' denotes the sonic point, $c_\mathrm{s}$ is the isothermal sound speed, and $ \rho_\mathrm{s}$ is the gas density at the sonic point. The location of the sonic point $r_\mathrm{s}$ is defined as
\begin{equation}
    2c^2_\mathrm{s} = \frac{G M_\mathrm{p}}{r_\mathrm{s}} - \frac{3GM_\star r^2_\mathrm{s}}{a^3},
      \label{eq:sound-point}
\end{equation}
where $M_\mathrm{p}$ is the planetary mass, $M_\star$ is the mass of the host star, $a$ is the orbital radius, and $G$ is the gravitational constant. If the penetration depth of stellar XUV photons in the atmosphere is located above the sonic point, we assume that $r_\mathrm{s} = R_\mathrm{XUV}$, where $R_\mathrm{XUV}$ is the location at which the H$_2$/He atmosphere becomes optically thick to stellar XUV photons. This situation occurs on highly inflated planets. The detailed modeling of the flow pattern of escaping hydrogen in such a scenario is beyond the scope of this study.
At a high UV flux, the effect of radiation recombination overwhelms the advection of hydrogen ions. In ionization equilibrium, if we neglect the effect of H$_3^+$ ion cooling, a balance between radiation recombination and photoionization at the photoionization base (i.e., the UV optical depth is unity) approximately yields
\begin{equation}
    \frac{F_\mathrm{UV}}{h\nu_0} \sigma_{\nu_0} n_0 \sim n^2_+ \alpha_\mathrm{rc},
      \label{eq:number-density}
\end{equation}
where $n_0$ and $n_+$ are the number density of neutral hydrogen atoms and hydrogen ions at the base, respectively, the recombination coefficient for hydrogen ions \citep{1995MNRAS.272...41S} is $\alpha_\mathrm{rc} = 2.7 \times 10^{-13} (T/10^4\,\mathrm{K})^{-0.9} \,\mathrm{cm}^3 \mathrm{s}^{-1}$, the photoionization cross-section of hydrogen is $\sigma_{\nu_0} = 6 \times 10^{-18} (h\nu_0/13.6\mathrm{eV})^{-3}\,\mathrm{cm}^2$ \citep[e.g.,][]{1978ppim.book.....S}, $F_\mathrm{UV}$ is the UV flux, and $T$ is the temperature. The number density of neutral hydrogen atoms at the base is $n_0 \sim 1/(\sigma_{\nu_0} H)$ using the scale height $H \sim kT/(g m_\mathrm{H}/2)$, where $k$ is the Boltzmann constant, $g$ is the gravitational acceleration, and $m_\mathrm{H}$ is the mass of the hydrogen atom. The density at the sonic point can be related to that at the photoionization base in the isothermal, hydrostatic structure:
\begin{equation}
    \rho_\mathrm{s} = \rho_\mathrm{base} \exp \left[-\frac{GM_\mathrm{p}}{c^2_\mathrm{s}} (R^{-1}_\mathrm{XUV} - r^{-1}_\mathrm{s})
        + \frac{3GM_\star}{2a^3c^2_\mathrm{s}} (r^2_\mathrm{s} - R^2_\mathrm{XUV})\right], 
      \label{eq:density_sonic}
\end{equation}
where $\rho_\mathrm{base}$ is the density at the photoionization base ($R_\mathrm{XUV}$).
This study assumed that an incoming UV photon ($h\nu_0$) has a characteristic energy of 20\,eV, which results in $T \sim 10000$\,K in the upper layer \citep{2009ApJ...693...23M}.

If stellar UV flux is low, a fraction of the hydrogen atoms in the upper atmosphere are ionized by UV photons. The gas temperature in the EUV-heated region is mainly determined by the balance between UV heating and cooling due to adiabatic expansion. In the EUV-driven regime, energy-limited hydrodynamic escape becomes the dominant mechanism of atmospheric loss from a planet.
Three-dimensional simulations of an outflow from a hot Jupiter \citep{2015ApJ...808..173T} suggest that the average mass loss rates are similar to those predicted in one-dimensional models.
The one-dimensional hydrodynamic mass loss rate of a H$_2$/He atmosphere of a planet $\dot{M}_\mathrm{p}$ is given by
\begin{equation}
  \dot{M_\mathrm{p}} = - \eta \frac{{R_\mathrm{p}}^3 L_\mathrm{XUV}(t)}{4  G  M_\mathrm{p}  a^2  K_\mathrm{tide}(R_\mathrm{RLO}/R_\mathrm{p})},
  \label{eq:mass-loss}
\end{equation}
$\eta$ is the heating efficiency due to stellar XUV irradiation, $L_\mathrm{XUV}$ is the stellar XUV luminosity, $R_\mathrm{p}$ is the planetary radius, and the Roche lobe radius is $R_\mathrm{RLO} \approx (M_\mathrm{p}/3 M_\star)^{1/3} a$ \citep{2007A&A...472..329E}. Since the heating efficiency for hydrogen-rich upper atmospheres was lower than 20\,\% \citep{2014A&A...571A..94S,2015SoSyR..49..339I}, we adopted $\eta = 0.1$. 
$K_\mathrm{tide}$ is the potential energy reduction factor owing to the effect of stellar tidal forces:
\begin{equation}
  K_\mathrm{tide}(\xi) = 1 - \frac{3}{2\xi} + \frac{1}{2\xi^3} < 1,\,\,\,\,\,\xi = \frac{R_\mathrm{RLO}}{R_\mathrm{p}}.
  \label{eq:K_tide}
\end{equation}

\subsection{The Interior Structure of a Planet}

The radius of a close-in planet with a H$_2$/He atmosphere is calculated by its interior structure in hydrostatic equilibrium under time-dependent stellar radiation \citep[see][]{2020ApJ...889...77H}. The planetary temperature is given by the equilibrium temperature: $T_\mathrm{eq} = \left[(1 - A) L_\star/(16\pi\sigma a^2)\right]^{1/4}$.
$T_\mathrm{eq}$ is the equilibrium temperature of a planet at its orbital radius, $L_\star$ is the time-dependent stellar luminosity, $\sigma$ is the Stefan-Boltzmann constant, and $A$ is the Bond albedo of the planet. For simplicity, $A$ was assumed to be zero in this study.
The rocky core inside a close-in planet consists of a silicate mantle and an iron core. The silicate mantle was described by the 3rd-order Birch-Murnagham EoS for MgSiO$_3$ perovskite \citep{2000PhRvB..6214750K,2007ApJ...669.1279S}. The Vinet EoS for $\epsilon$-Fe \citep{2001GeoRL..28..399A} was used for the iron material in the core. The Thomas-Fermi Dirac EoS \citep{1967PhRv..158..876S} was applied to high-pressure EoS for MgSiO$_3$ at $P \geq 4.90$\,TPa and Fe at $P \geq 2.09 \times 10^4$\,GPa \citep{2007ApJ...669.1279S,2013PASP..125..227Z}.
The iron material in the rocky core was 30\,\% of the core mass.
The pressure-temperature profile in the H$_2$/He envelope was calculated using the SCvH EoS \citep{1995ApJS...99..713S}.
This study assumed that $R_\mathrm{p}(t) = R_\mathrm{XUV}$ for close-in super-Earths, where $R_\mathrm{p}(t)$ is the planetary radius at a given time $t$ and $R_\mathrm{XUV}$ is the location of the photoionization base. We define the planetary radius as the photosphere, which is given by $R_\mathrm{p}(t) = R_\mathrm{bc} + R_\mathrm{atm}$, where $R_\mathrm{bc}$ is the radiative–convective boundary derived from the interior modeling and $R_\mathrm{atm}$ is the photospheric correction given in \citet{2014ApJ...792....1L}.
The thermal evolution of a super-Earth with a H$_2$/He atmosphere is computed by calculating its interior structure at each timestep, followed by the mass loss and subsequent orbital evolution. The atmospheric mass loss rate of a super-Earth is determined by equation (\ref{eq:mass-loss}), using $R_\mathrm{p}(t)$.
If $R_\mathrm{RLO}$ is smaller than $R_\mathrm{p}$, a planet can no longer gravitationally bind its upper atmosphere, leading to the onset of Roche lobe overflow. We considered the initial H$_2$/He atmosphere of a planet that satisfies $R_\mathrm{p} < R_\mathrm{RLO}$.
In this study, the transition point from the radiation recombination-limited to the energy-limited regime was determined by calculating the mass loss rate of each regime and adopting the regime yielding a smaller rate (see Appendix \ref{appendix:transition}).

\subsection{Orbital Evolution of an Evaporating Planet} \label{subsec:orbital_evolution}

We modeled the orbital evolution of a planet by photoevaporation based on \citet{2016CeMDA.126..227J}'s approach for the orbital evolution of a close-in gas giant driven by Roche lobe overflow.
Although \citet{2016CeMDA.126..227J} assumed that the planet is in a circular orbit around its host star, we consider that a star and a planet have a Keplerian circular orbit around their common center of mass (i.e., the barycenter of the star-planet system). Note that the non-zero eccentricity of a planet has a negligible effect on the orbital evolution in our simulations (see Section \ref{sec:discussions}). We considered the isotropic mass loss from a planet in this study. The effects of the anisotropic mass loss are also discussed in Section \ref{sec:discussions}.

The orbital angular momentum of the star-planet system $L$ is given by
\begin{equation}
    L = M_\star {a_\star}^2 \Omega + M_\mathrm{p} {a_\mathrm{p}}^2 \Omega = M_\star M_\mathrm{p} \sqrt{\frac{Ga}{M}},
    \label{eq:L}
\end{equation}
and the time derivative of $L$ is given by
\begin{equation}
    \frac{\dot{L}}{L} = \frac{\dot{M_\star}}{M_\star} + \frac{\dot{M_\mathrm{p}}}{M_\mathrm{p}} + \frac{1}{2} \frac{\dot{a}}{a} - \frac{1}{2} \frac{\dot{M}}{M},
    \label{eq:Ldot}
\end{equation}
$\Omega$ is the Keplerian angular velocity, $M_\star$ is the stellar mass, and $M$ is the total mass of the star-planet system (i.e., $M = M_\star + M_\mathrm{p}$), $a$ (= $a_\star + a_\mathrm{p}$) is the orbital separation between the planet and star, and $a_\star$ and $a_\mathrm{p}$ are the orbital distances of the star and the planet from the barycenter of the system, respectively.
If $\dot{L} = 0$ and $\dot{M}_\star = -\dot{M}_\mathrm{p}$, we obtain $\dot{a}/a = - 2\dot{M}_\mathrm{p}(M_\star-M_\mathrm{p})/(M_\star M_\mathrm{p}) > 0$, which corresponds to the conservative mass transfer from a planet to its host star.
On the other hand, when the material escaping from a planet carries away its orbital angular momentum, that is $\dot{L} = \dot{M_\mathrm{p}}a^2_\mathrm{p}\Omega$, equation (\ref{eq:Ldot}) yields $\dot{a}/a = - \dot{M_\mathrm{p}}/M > 0$.
Thus, the mass loss from a planet via atmospheric escape causes orbital expansion \citep[see also][]{1976ApJ...204..879A,1984MNRAS.208..763L}. 

The change in the orbital angular momentum of the system determines the orbital migration of a planet.
The change in the orbital angular momentum of the system depends on the dynamics of an evaporative wind, such as the velocity of the atmospheric particles and the direction of the flow.
The detailed behavior of a planetary wind launched by stellar XUV radiation, however, is beyond the scope of this study.
Instead, the change rate in the angular momentum of a star-planet system is defined as $\dot{L} = (1-\chi)\dot{M_\mathrm{p}} {a_\mathrm{p}}^2 \Omega$, where $\chi$ ($0 \leq \chi \leq 1$) is the fraction of the angular momentum conserved in the system \footnote{Note that $\chi$ nearly corresponds to $1 - \gamma\delta$ in \cite{2016CeMDA.126..227J}.}. Substituting this into equation (\ref{eq:Ldot}), we obtain
\begin{equation}
  \frac{\dot{a}}{a} = -2 \chi \dot{M_\mathrm{p}} \frac{M_\star - M_\mathrm{p}}{M_\star M_\mathrm{p}}  - (1-\chi)\frac{\dot{M_\mathrm{p}}}{M} > 0.
  \label{eq:orbital_evolution}
\end{equation}

Although $\chi$ may vary with time due to the geometry of the evaporative wind, $\chi$ is assumed to be constant in this study. The evaporative wind from a planet would produce a cometary tail or torus of a fast H I atom due to charge exchange between the H I atom escaping from a planet and a proton in a stellar wind \citep[e.g.,][]{2016ApJ...832..173S,2020MNRAS.493.1292D}, however, the detailed behavior and geometry of the escaping atmosphere is also not considered in this study.
Note that the escaping atmospheric mass of $- \chi \dot{M_\mathrm{p}}$, which is conserved in the star-planet system, does not necessarily have to accrete onto the star; what is important for the orbital evolution is not the mass accretion onto the star but whether the escaping mass is bound by the system just after it flows out of the planet.
If a star-planet system conserves part of the orbital angular momentum of an escaping atmosphere, the system needs to compensate for the angular momentum by the orbital expansion of a planet.
On the other hand, the change in the stellar mass has almost no effect on the orbital evolution because $M_\star \gg M_\mathrm{p}$, and $a_\star$ is too small to significantly increase the orbital angular momentum of the star. Thus, we simply used $\dot{M_\star} = - \chi \dot{M_\mathrm{p}}$\footnote{\cite{2016CeMDA.126..227J} assumed no mass accretion onto a host star ($M_\star = 0$).} for equation (\ref{eq:Ldot}) to eliminate $\dot{M_\star}$ to obtain equation (\ref{eq:orbital_evolution}). This simplified treatment would hardly affect our results.

\section{RESULTS} \label{sec:results}

\subsection{Orbital Evolution of an Evaporating Super-Earth} \label{subsec:results_single}

\begin{figure*}[ht!]
\centering
\includegraphics[width=\linewidth,clip]{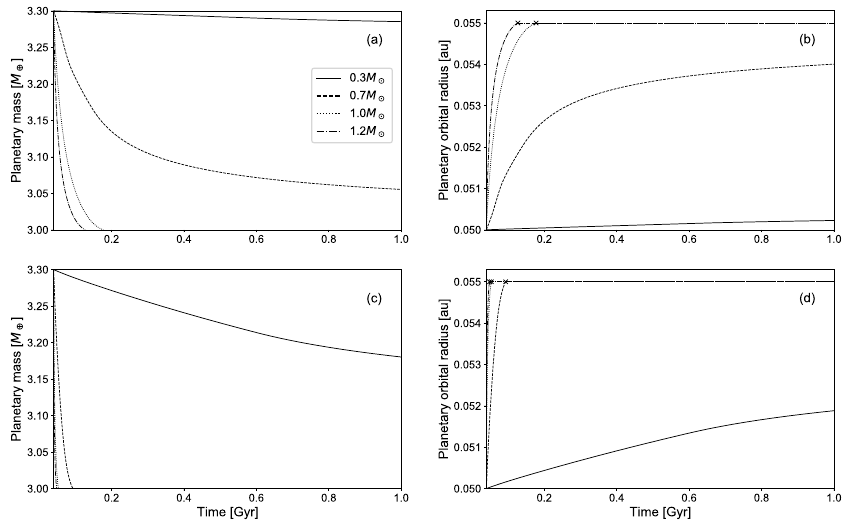}
\caption{Time evolution of the mass and orbital radius of a close-in super-Earth that experiences the atmospheric escape for 1\,Gyr after disk dispersal. A super-Earth with a rocky core of 3\,$M_\oplus$ was initially located at 0.05\,au around FGKM-type stars: 0.3\,$M_\odot$ (solid lines), 0.7\,$M_\odot$ (dashed lines), 1.0\,$M_\odot$ (dotted lines), and 1.2\,$M_\odot$ (dot-dashed lines). The initial mass fraction of the H$_2$/He atmosphere was 10\,\% of the core mass. $\chi$ is assumed to be 0.5.
Crosses represent the point where the H$_2$/He atmosphere is completely lost. The bottom panels show the results of the case where the stellar XUV luminosity is ten times higher than that of our standard model.}
\label{fig:stellar_dependence}
\end{figure*}

Figure \ref{fig:stellar_dependence} demonstrates the mass loss of a close-in super-Earth around FGKM-type stars and subsequently, its orbital migration. A super-Earth initially has a rocky core of 3\,$M_\oplus$ surrounded by a H$_2$/He atmosphere of 10\,\% relative to the core mass. A planet continues to move outward from 0.05\,au while losing its atmosphere.
Once the atmosphere of a planet is completely lost or the stellar XUV flux decreases significantly (typically, $t \gtrsim \tau_\mathrm{sat}$), the outward migration ceases. The increase in the orbital radius of an evaporating super-Earth is determined by atmospheric mass loss caused by photoevaporation. The evaporative loss of a 10\,wt\% H$_2$/He atmosphere from a planet results in an increase in its orbital radius by approximately 10\,\% (see also equation (\ref{eq:orbital_evolution})).

The mass loss rate from a planet is controlled by the strength of the stellar XUV flux. A high stellar luminosity also enhanced the inflation of the planetary atmosphere. The more massive the star a planet orbits around is, the more violent and rapid atmospheric escape occurs due to intense stellar irradiation. As a planet moves outward rapidly, the incident stellar XUV flux quickly decreases as $F_\mathrm{XUV} \propto a^{-2}$. The integrated XUV flux that a planet receives for $\sim 1$\,Gyr determines the atmospheric mass loss from itself. 
As shown in the upper panels of Figure \ref{fig:stellar_dependence}, planets around FG-type stars rapidly lose their entire atmospheres, while for those around K-type stars, the mass loss rate decreases significantly after the stellar XUV flux starts to decrease at $\sim 0.1$\,Gyr.
A planet around an M-type star maintains the initial atmosphere over $\sim 1$\,Gyr because of a lower XUV flux.
We also show a case where the stellar XUV luminosity is ten times as high as our standard model (Figures \ref{fig:stellar_dependence}c and \ref{fig:stellar_dependence}d) because of its large uncertainty. In a high-XUV model, planets around K-type and FG-type stars rapidly lose their entire atmospheres. Even a planet around an M-type star loses one-third of its initial atmosphere in 1\,Gyr.
Hereafter, results are observed in standard XUV models.

\begin{figure*}[ht!]
\centering
\includegraphics[width=\linewidth,clip]{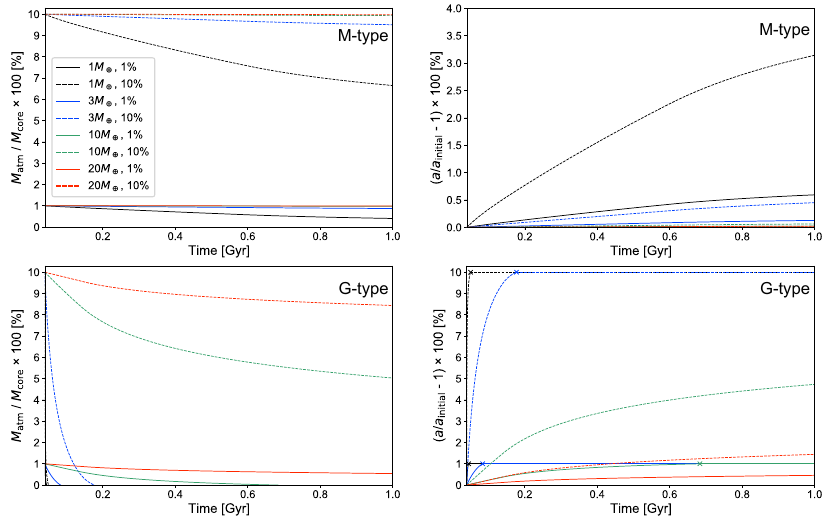}
\caption{Time evolution of the atmospheric mass fraction and the orbital radius of super-Earths with four different core masses for 1\,Gyr around an M-type star with 0.3\,$M_\odot$ (top panels) and a G-type star with 1.0\,$M_\odot$ (bottom ones): 1\,$M_\oplus$ (black lines), 3\,$M_\oplus$ (blue ones), 10\,$M_\oplus$ (green ones), and 20\,$M_\oplus$ (red ones).
The initial mass fraction of the H$_2$/He atmosphere was 1\,\% (solid lines) and 10\,\% (dashed lines) of the core mass.
Both the initial location of a planet and $\chi$ are the same as those in Figure \ref{fig:stellar_dependence}.
$M_\mathrm{atm}$ is the mass of the H$_2$/He atmosphere, and $a_\mathrm{initial}$ is the initial location of the planet.
Crosses represent the point that a H$_2$/He atmosphere is completely lost.}
\label{fig:core_dependence}
\end{figure*}

Figure \ref{fig:core_dependence} shows the change rate in the atmospheric mass and the orbital radius of super-Earths with different core masses around a G-type (1.0\,$M_\odot$) and M-type star (0.3\,$M_\odot$). Less massive cores are prone to lose a H$_2$/He atmosphere by photoevaporation because the inflation effect makes their upper atmospheres loosely bound. The existence of a massive core helps the retention of a H$_2$/He atmosphere.
Small planets with a rocky core of $\lesssim 3\,M_\oplus$ around a G-type star completely lose their atmosphere in $\lesssim 0.1$\,Gyr, regardless of the initial atmospheric mass. Massive cores with $\geq 10\,M_\oplus$ can retain their atmospheres over 1\,Gyr, especially if they have a substantial amount of the initial atmosphere.

\begin{figure*}[ht!]
\centering
\includegraphics[width=\linewidth,clip]{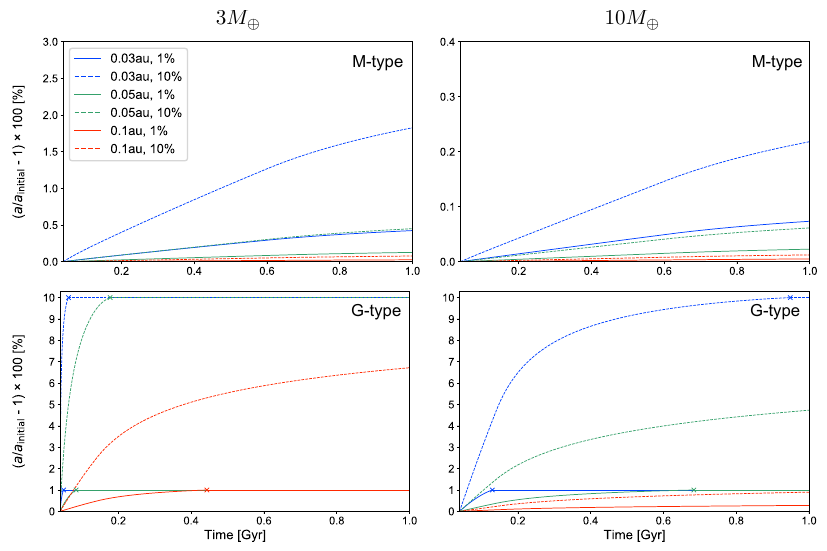}
\caption{Same as the right panels in Figure \ref{fig:core_dependence}, but for different initial locations: 0.03\,au (blue lines), 0.05\,au (green lines), and 0.1\,au (red lines). The upper panels show the orbital evolution of a 3\,$M_\oplus$ (left panels) and 10\,$M_\oplus$ super-Earth (right panels) around an M-type star (0.3\,$M_\odot$) that initially has a H$_2$/He atmosphere of 1\,wt\% (solid lines) and 10\,wt\% (dashed lines) of the core mass, respectively. The lower panels are similar to the upper panels, but for super-Earths around a G-type star (1.0\,$M_\odot$). $\chi$ is assumed to be 0.5.
Crosses represent the point that a H$_2$/He atmosphere is completely lost.}
\label{fig:orbital_radius_dependence}
\end{figure*}

As a planet is initially farther away from its host star, the H$_2$/He atmosphere is likely to survive photoevaporation for 1\,Gyr. 
Figure \ref{fig:orbital_radius_dependence} shows the orbital evolution of super-Earths with different initial locations.
A close-in super-Earth with $\geq 0.03$\,au around an M-type star avoids the complete loss of H$_2$/He atmosphere.
The outward migration of a super-Earth around an M-type star driven by its atmospheric escape is negligible unless it has a small core of $\leq 3\,M_\oplus$ with a massive H$_2$/He atmosphere at $\leq 0.03$\,au.
The critical orbital radius at which a super-Earth around an M-type star loses its entire atmosphere is $\sim$ 0.01--0.02\,au.
This suggests that an ultra-short period (USP) planet around an M-type star should have no atmosphere unless they have either a massive core or a substantial amount of the initial atmosphere, which is consistent with the mass-radius relation of observed USP planets (USPs) (see Figure 20 in \citet{2021AJ....162..161H}).
A G-type star blows off most of the H$_2$/He atmosphere that a super-Earth initially possesses.
Because a close-in super-Earth with a 1\,wt\% (10\,wt\%) H$_2$/He atmosphere at $\lesssim 0.03$\,au around a G-type star completely loses the initial H$_2$/He atmosphere by photoevaporation, its orbital radius increases by 1\,\% (10\,\%) of its initial location. This implies that the observed close-in super-Earths around a G-type star might have migrated outward for $\sim 1$\,Gyr if they had a primordial atmosphere. A super-Earth with a massive core of $\geq 10\,M_\oplus$ and a massive H$_2$/He atmosphere at $> 0.1$\,au around a G-type star can maintain almost all of the initial atmosphere at its current location. 

\begin{figure}[ht!]
\centering
\includegraphics[width=\linewidth,clip]{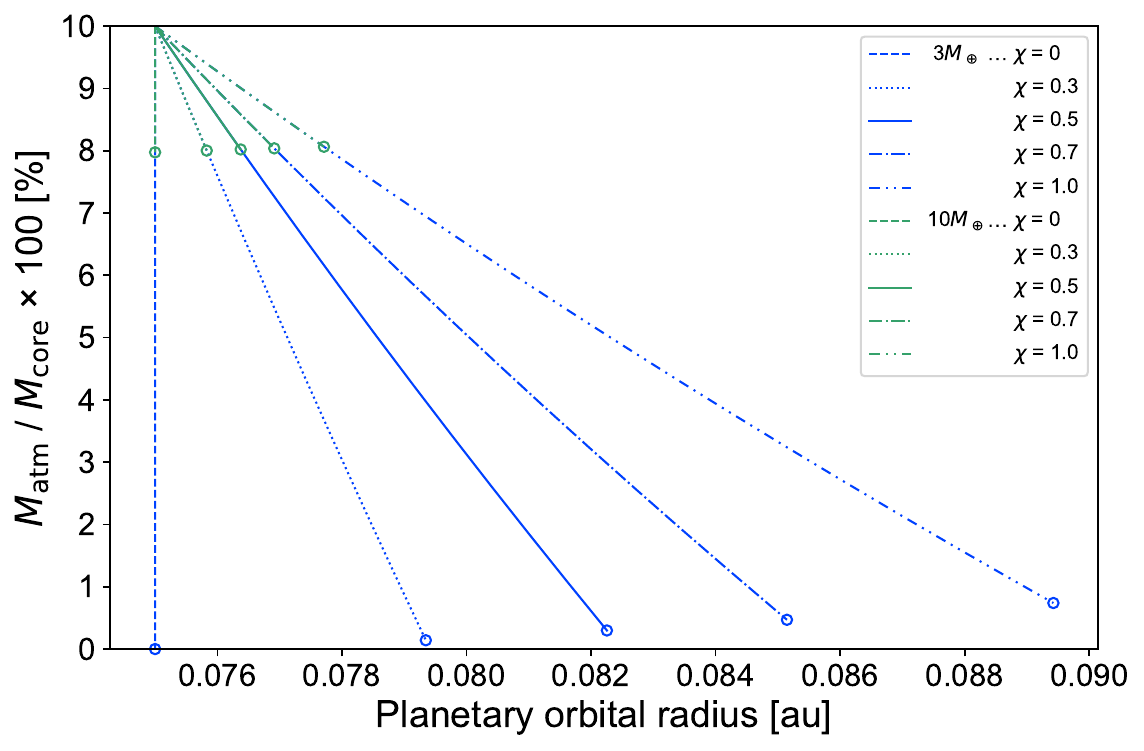}
\caption{Atmospheric and orbital evolution of super-Earths with rocky cores of 3\,$M_\oplus$ (blue lines) and 10\,$M_\oplus$ (green lines) around a G-type star (1.0\,$M_\odot$). Each super-Earth initially has a H$_2$/He atmosphere of 10\,\% of its core mass at 0.075\,au.
Each line style corresponds to a different $\chi$ value:0 (dashed lines), 0.3 (dotted lines), 0.5 (solid lines), 0.7 (dot-dashed lines), and 1.0 (dot-dot-dashed lines), respectively. The open circles denote the final locations of each planet.}
\label{fig:chi_dependence}
\end{figure}

Figure \ref{fig:chi_dependence} shows the orbital evolution of a 3\,$M_\oplus$ and 10\,$M_\oplus$ super-Earth with five different $\chi$ values ($\chi$ = 0, 0.3, 0.5, 0.7, and 1). We found that the larger $\chi$ is, the further the planet migrates.
A star-planet system conserves part of the orbital angular momentum of an escaping atmosphere. Because atmospheric escape reduces the total mass of a planet, the orbital expansion of a planet compensates for the mass loss to satisfy the conservation of angular momentum. The fraction of the leftover orbital angular momentum, which is defined by $\chi$, regulates the orbital migration of a planet (see also Section \ref{subsec:orbital_evolution}). Thus, a large $\chi$ results in an efficient outward migration of the planet.
Two super-Earths with a given $\chi$ value follow the same path on the atmospheric mass fraction-orbital radius plane, as predicted from equation (\ref{eq:orbital_evolution}). However, the migration rate of a planet strongly depends on its initial conditions; e.g., the core mass and orbital radius. As a planet migrates outward, it should undergo a less violent atmospheric escape. As a result, a different $\chi$ value affects the final atmospheric mass of a planet. 
The net angular momentum loss by the atmospheric escape plays a key role in accurately determining the orbital evolution and mass loss process of a close-in super-Earth.

\subsection{Period-Radius Relationship of Close-in Super-Earths}

We demonstrate the period-radius distribution of planets around a Sun-like star (an FGK-type star) and an M-type star. We simulate the orbital evolution of an evaporating super-Earth in $\sim 10^5$ single-planet systems for each type of star.
The mass of an FGK-type star follows a Gaussian distribution with a mean of 1\,$M_\odot$ and a standard deviation of 0.15\,$M_\odot$.
The orbital period distribution of a planet between 1\,day and 100\,days is given by the {\it Kepler} planet sample, following the population synthesis model in \citet{2017ApJ...847...29O}:
\begin{equation}
  \frac{dN}{d\log P} \propto \left\{
  \begin{array}{ll}
    \mathrm{constant} & \,\,\,\, \mathrm{for}\,\, P > 7.6 \, \mathrm{days},\\
    P^{1.9} & \,\,\,\, \mathrm{for}\,\, P \leq 7.6 \, \mathrm{days}.
  \end{array}
  \right.
  \label{eq:period_dis}
\end{equation}

A close-in super-Earth has a rocky core whose mass distribution takes the following Rayleigh distribution:
\begin{equation}
  \frac{dN}{dM_\mathrm{core}} \propto M_\mathrm{core}\exp\left(-\frac{M_\mathrm{core}^2}{2\sigma_\mathrm{M}^2}\right),
  \label{eq:core_Rayleigh}
\end{equation}
where $M_\mathrm{core}$ is the core mass, and $\sigma_\mathrm{M}$ is set to 3\,$M_\oplus$.
The initial atmospheric mass fraction of a planet is given by a log-uniform distribution in the range of 1\,\% to 10\,\% with respect to its core mass. The $\chi$ value was set to $0.5$.

\begin{figure*}[ht!]
\centering
\includegraphics[width=\linewidth,clip]{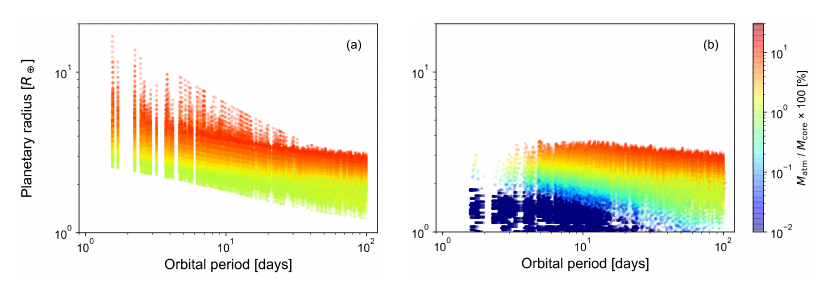}
\caption{Orbital period-planetary radius distribution of planets around G-type stars (a) in the initial state and (b) after 1\,Gyr. The color contour shows the atmospheric mass fraction of the planet.}
\label{fig:scatter_initial_final}
\end{figure*}

\begin{figure*}[ht!]
\centering
\includegraphics[width=0.9\linewidth,clip]{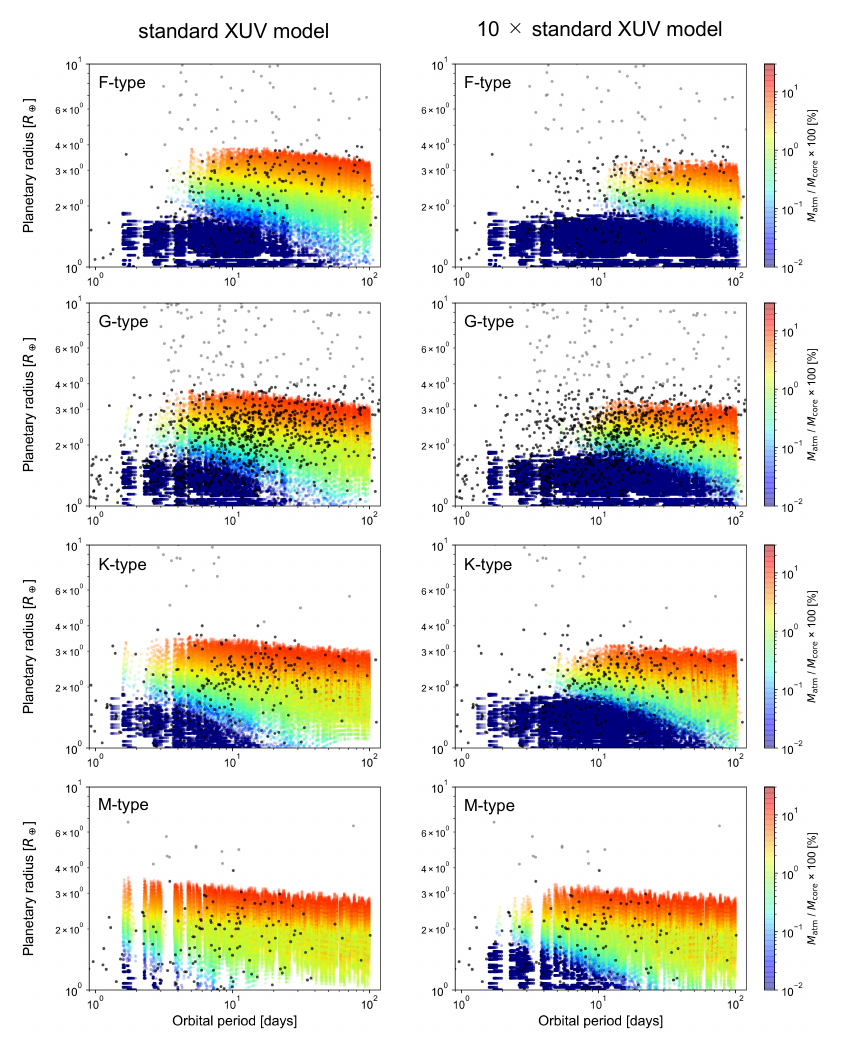}
\caption{Period-radius distribution of the planets around F-type stars (1.1--1.4\,$M_\odot$), G-type stars (0.8--1.0\,$M_\odot$), K-type stars (0.6--0.7\,$M_\odot$), and an M-type stars (0.2--0.5\,$M_\odot$) after the atmospheric escape with $\chi = 0.5$ for 1\,Gyr. The detected planets with $\leq 4\,R_\oplus$ and $> 4\,R_\oplus$ are plotted as black and gray filled circles, respectively. The left panels show the results of our standard XUV model, and the right panels show results in the case where a stellar XUV flux is ten times as strong as the standard model.}
\label{fig:scatter}
\end{figure*}

Figure \ref{fig:scatter_initial_final} shows the orbital period distribution of planets around G-type stars (0.8--1.0\,$M_\odot$) in the initial state and after the atmospheric escape for 1\,Gyr. 
Many super-Earths having no atmospheres appear at $\lesssim 30$\,days.  
Close-in super-Earths with $P \lesssim 10$\,days lose a large fraction of H$_2$/He atmosphere and move outward. As a result, there is a lack of close-in super-Earths with radii larger than 2\,$R_\oplus$ after 1\,Gyr.

The final period-radius distributions of planets around FGKM-type stars are shown in Figure \ref{fig:scatter}. We simulated the orbital evolution of planets in two XUV flux models because of uncertainties in the stellar XUV flux: (left) the standard XUV model, and (right) a 10$\times$ standard XUV model.
As stellar XUV flux increases with increasing stellar mass, close-in planets around a massive star are more vulnerable to photoevaporative mass loss. Close-in super-Earths with $P \lesssim$ 3--5\,days around FGK-type stars lose almost all the H$_2$/He atmospheres and then move outward, whereas mass loss processes are less efficient for those around M-type stars.
Small planets with $\lesssim 2\,R_\oplus$ are bare rocky cores.
Detected planets with $\leq 4\,R_\oplus$ and $> 4\,R_\oplus$ around the FGKM-type stars are also represented by black and gray filled circles in Figure \ref{fig:scatter} \footnote{The data come from NASA Exoplanet Archive (\url{https://exoplanetarchive.ipac.caltech.edu}).}. The period-radius distributions of simulated planets agree with observations, except for those of planets with $P \lesssim 10$\,days with 2--4\,$R_\oplus$ around FG-type stars in a high-XUV flux model.
We can see the Neptune desert in the planet population around the FGK-type stars.
Note that there are no USPs at $\lesssim 1$\,day and large planets with $\gtrsim 4\,R_\oplus$ in our simulations because of the initial distributions of planets.

\subsection{Occurrence Rates of Close-in Super-Earths}

\begin{figure*}[ht!]
\centering
\includegraphics[width=\linewidth,clip]{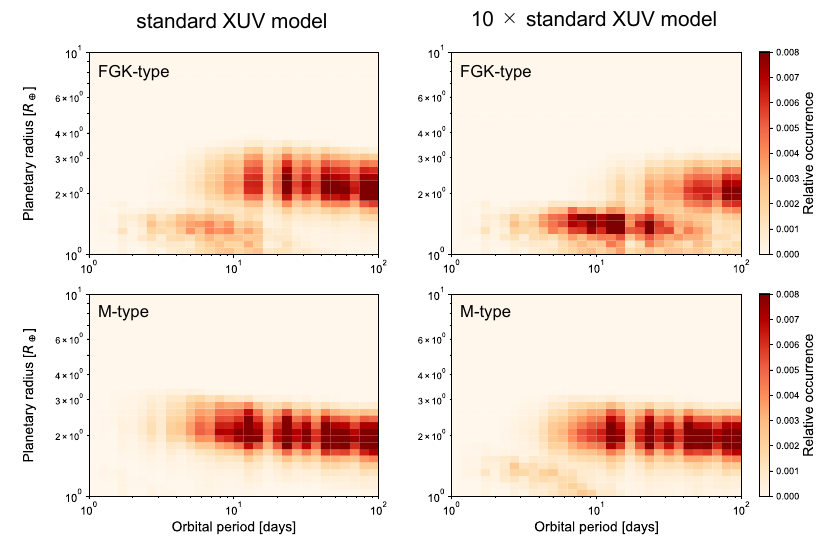}
\caption{Period-radius distribution of planets around FGK-type stars (Top: 0.6--1.4\,$M_\odot$) and M-type stars (Bottom: 0.2--0.5\,$M_\odot$) after the atmospheric escape for 1\,Gyr. The left panels show the results of our standard XUV model, and the right panels show those in the ten times XUV flux model. The color contour represents the relative occurrence rate of the planets. The mass of an FGK-type star follows a Gaussian distribution with a mean of 1\,$M_\odot$ and a standard deviation of 0.15\,$M_\odot$.}
\label{fig:heat}
\end{figure*}

Figure \ref{fig:heat} shows period-radius distributions of planets around FGK-type stars and M-type stars after the atmospheric escape with $\chi = 0.5$ for 1\,Gyr.
We see the lack of planets with radii of 1.5--2.0\,$R_\oplus$ for $P \sim 10$\,days around FGK-type stars on the period-radius plane. As a high XUV flux enhances the mass loss from a planet at $\lesssim 10$\,days, small planets with $\lesssim 2\,R_\oplus$ are more common. On the other hand, there is no clear radius gap around M-type stars because of less efficient mass loss from close-in planets, although a high XUV model produces bare rocky planets at $\lesssim 10$\,days. These results suggest that the presence of a radius gap is sensitive to the intensity of the stellar XUV flux. 

\begin{figure*}[ht!]
\centering
\includegraphics[width=\linewidth,clip]{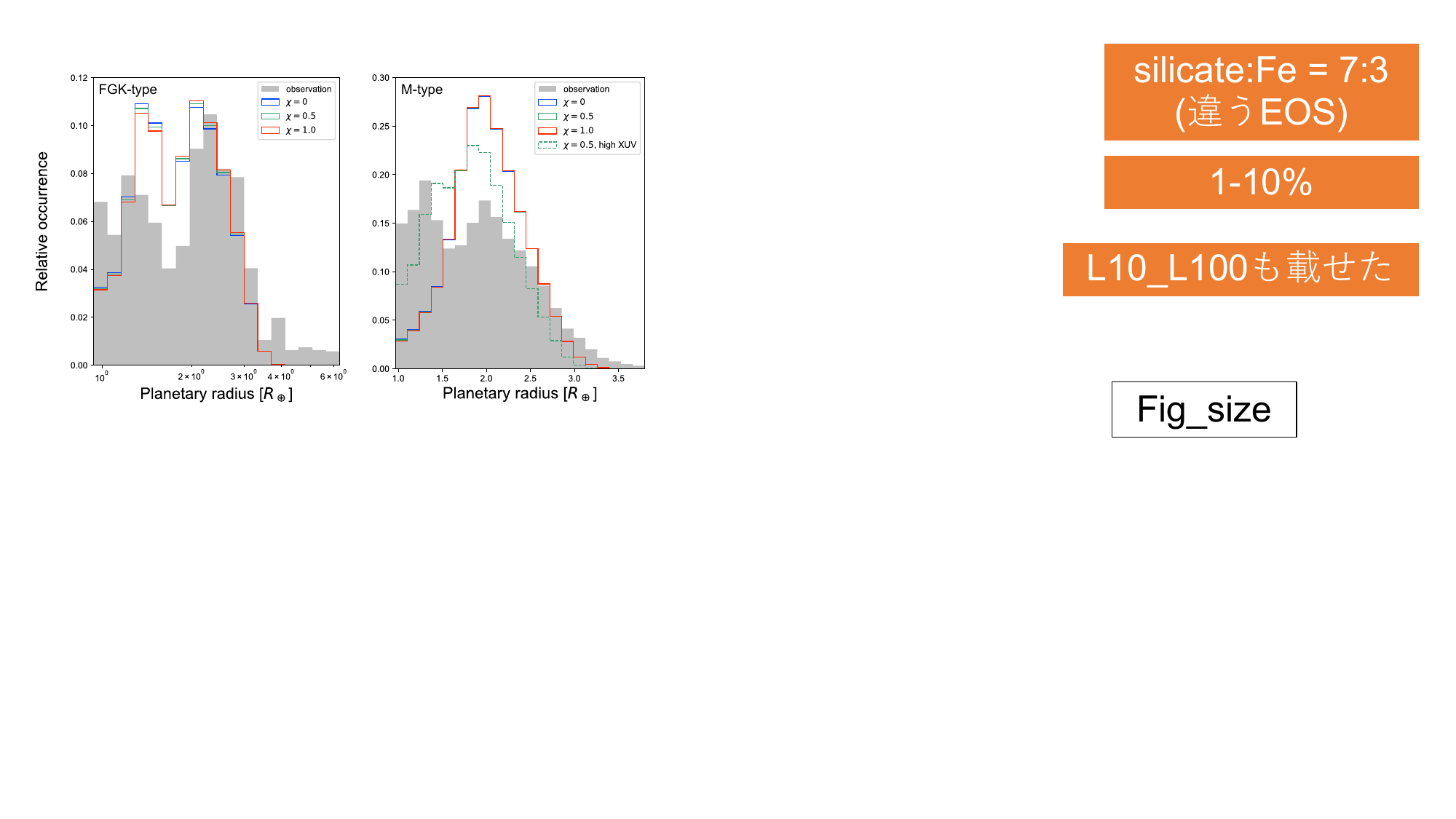}
\caption{Final size distribution of planets with $P < 100$\,days around FGK-type stars and M-type stars.
Gray histograms show the completeness-corrected distribution of planets with $P < 100$\,days around FGK-type stars \citep{2018AJ....156..264F}, and M-type stars \citep{2020AJ....159..211C}.
Each color histogram represents the weighted sum of the planet size distributions after 1\,Gyr in the two stellar XUV models for $\chi = 0$ (blue), 0.5 (green), and 1.0 (red). For the M-type star, the higher XUV case (weighted sum of 10$\times$ and 100$\times$ standard XUV) is also shown by the dashed line.}
\label{fig:size}
\end{figure*}

Figure \ref{fig:size} shows the weighted sum of planet size distributions in the two stellar XUV models (the standard XUV model and 10$\times$ standard XUV model) with $P < 100$\,days after 1\,Gyr for $\chi = 0$ (blue), 0.5 (green), and 1.0 (red).
Gray histograms show the completeness-corrected distribution of planets with $P < 100$\,days around FGK-type stars \citep{2018AJ....156..264F}, and M-type stars \citep{2020AJ....159..211C}.
We find that there is a radius gap near 1.6--1.8\,$R_\oplus$ in the bimodal size distribution of planets around FGK-type stars in our simulations.
The occurrence rate ratio of planets within and beyond a radius gap increases with $\chi$ because the outward migration of a planet can mitigate atmospheric escape.
The peak magnitudes and the location of a gap in the planet size distribution are sensitive to the composition of a rocky core (see Figure 8 in \citet{2017ApJ...847...29O}).
In contrast, we can see no clear deficit of planets with 1.5--2.0\,$R_\oplus$ around M-type stars.
Many planets around M-type stars retain their H$_2$/He atmospheres after 1\,Gyr in our simulation, while in observation there exists a lot of small planets which may have no atmospheres. Our results of M-type stars are not consistent with observation, which shows a radius gap near 1.5--2.0\,$R_\oplus$, because of their less intensive stellar XUV flux.
This implies that a young M-type star may have higher XUV luminosity. Indeed, we confirmed that a stronger XUV flux (dashed line for the M-type star in Figure \ref{fig:size}) can blow off the atmosphere of planets near 2\,$R_\oplus$ to increase the proportion of planets with smaller radii in the size distribution of planets around M-type stars.
Note that the size distribution of close-in super-Earths with a paucity in their occurrence rates changes with their initial atmospheric mass fractions, the distribution of their core masses, and the range of their orbital period.
Observations show that there exist a lot of small planets with $P \gtrsim 20$\,days (i.e., they had no primordial atmosphere), so in practice, we may need to set up initial distributions of planets which include bare rocky planets. For example, the initial distribution following the updated core-mass function and initial atmospheric mass function in \cite{2021MNRAS.503.1526R} would alter our results, especially for the radius gap; the radius gap can be reproduced even if we adopt a standard XUV model because there exist small planets without even atmospheric escape.

\section{DISCUSSIONS} \label{sec:discussions}

In this study, we considered the atmospheric escape from a planet as an isotropic ejection; that is, no torque caused by mass ejection acts on a planet. However, a torque driven by an anisotropic evaporative wind may have a non-negligible effect on the orbital evolution of a planet. If a planet loses its atmosphere anisotropically, the planet can move outward more significantly depending on the ejection direction \citep{2012A&A...537L...3B, 2015MNRAS.452.1743T}.
Detailed three-dimensional hydrodynamic simulations of atmospheric particles flowing out of a planet are essential for scrutinizing the geometry of the evaporative wind such as the ejection direction and the angular momentum loss by the evaporative wind.

Although $\chi$ was treated as a parameter in this study, our theoretical prediction of planetary mass-orbital distributions has the potential to better understand the dynamics of evaporative wind from future observations of close-in super-Earths. More super-Earth around M-type stars are expected to be discovered by ongoing high-precision Doppler surveys, such as CFHT/SPIRou, CARMENES, Subaru/IRD, VLT/ESPRESSO, Gemini/MAROON-X, and HET/HPF. Observed planet populations around FGKM-type stars help to retrieve information about the atmospheric escape and orbital evolution of close-in planets.

The mass loss from a planet increases the orbital radius by 1--10\,\% of its initial location.
The orbital expansion of a planet in a compact, multi-planet system, such as the TRAPPIST-1 and Kepler-186 systems, likely rearranges the orbital configurations of planets. Furthermore, the outward migration of a planet in a non-resonant system may trigger the trapping of outer planets into mean-motion resonance.
In addition, the mass loss processes of close-in planets in a multi-planet system may cause dynamical instability. Planetary systems are apt to undergo orbital instability with increasing multiplicity and mass loss of planets and their host star \citep{2020ApJ...893...43M}. 
On the other hand, \cite{2015MNRAS.452.1743T} suggests that outward migration driven by anisotropic evaporative wind would have pushed the systems deeper inside resonances.
Thus, the orbital evolution by the mass loss of close-in planets due to the photoevaporation would alter the fraction of resonant systems expected by the present simulation. The detailed orbital evolution of multiple planets due to the mass loss will be an important future work.

The tidal interaction between a close-in planet and its host star causes an orbital decay. As the orbit of a planet expands because of its atmospheric mass loss, the strength of tidal interactions decreases quickly as $\tau_\mathrm{tide} (= a_\mathrm{p}/\dot{a_\mathrm{p}}) \propto a^{13/2}_\mathrm{p}$ (see Appendix \ref{appendix:tide} for more details). 
For a close-in super-Earth around a main-sequence star, the mass loss should be experienced during a high stellar XUV flux ($t \lesssim 0.1$\,Gyr). The tidal damping of a planet's orbit occurs at $> 1$\,Gyr. As a result, the inward movement of an evaporating super-Earth due to tidal interactions has a negligible effect on its orbital expansion due to mass loss. In this study, we neglect the tidal effects on the orbital evolution of a close-in super-Earth. Following \citet{2009ApJ...705L..81V}, we simulated the orbital evolution of a close-in super-Earth that undergoes mass loss, including tidal effects, and confirmed the validity of our assumption (see also Appendix \ref{appendix:tide}).

This study considered a close-in planet in a circular orbit. The non-zero eccentricity of a planet causes time variations in stellar XUV irradiation, which are synchronized with orbital motion. 
The tidal circularization timescale for a close-in super-Earth is $> 1$\,Gyr (see equation (\ref{eq:tide-dissipation}) in the Appendix). As shown in Figures \ref{fig:stellar_dependence}–\ref{fig:orbital_radius_dependence}, a planet loses its H$_2$/He atmosphere during a high XUV phase ($t \lesssim 0.1$\,Gyr.
An initially eccentric super-Earth may have non-zero eccentricity while losing its atmosphere.
An eccentric planet receives the stellar flux averaged over the entire orbit of $\langle F_\mathrm{XUV} \rangle = F_\mathrm{XUV}/\sqrt{1-e^2}$. For $e \lesssim 0.1$, as predicted by the formation of close-in super-Earths via giant impacts \citep[e.g.,][]{2017AJ....154...27M}, the average stellar flux is greater by only $< 0.5$\,\%, compared to a planet in a circular orbit. The eccentric orbit of a planet decays rapidly owing to the tidal interactions. The eccentricity of a planet damps faster than its orbital radius if $0 < e \ll 1$ (see also equation (\ref{eq:tide-dissipation})).
Consequently, the orbit of an eccentric planet decays in a similar fashion to a planet in a circular orbit. Therefore, the non-zero eccentricity of a planet has a negligible effect on mass-loss-driven orbital evolution in our simulations.

\section{SUMMARY} \label{sec:summary}

We examined the effect of atmospheric escape driven by stellar X-ray and UV irradiation on the orbital evolution of close-in super-Earths around FGKM-type stars.
This study was the first comprehensive attempt to evaluate the orbital migration of close-in super-Earths via hydrodynamic escape.
The orbital expansion of a close-in planet compensates for the mass loss to satisfy the conservation of the orbital angular momentum in the system.
The less angular momentum the escaping atmospheric material carries away from the system, the further the planet migrates.
The rate of increase in the orbital radius of an evaporating planet is approximately proportional to that of the atmospheric mass loss during a high stellar XUV phase.
The more massive the star a planet orbits around is, the more violent and rapid atmospheric escape occurs due to intense stellar irradiation.
Less massive cores are more vulnerable to photoevaporative mass loss because the inflation effect loosely binds their upper atmospheres.
Small planets around massive stars are likely to experience mass-loss-driven orbital migration.

Super-Earths with a rocky core of $\lesssim 10\,M_\oplus$ and a H$_2$/He atmosphere at $\lesssim$ 0.03--0.1\,au ($\lesssim$ 0.01--0.03\,au) around G-type stars (M-type stars) are prone to the outward migration driven by photoevaporation.
Since USPs around an FGK-type star lose most of their atmospheres in 1\,Gyr, they should have no atmospheres. On the other hand, USPs around an M-type star have the possibility of retaining a few atmosphere because of a less intensive stellar XUV flux in case that a planet has either a massive core or a substantial amount of the initial atmosphere.

The observed close-in super-Earths around FGKM-type stars should have experienced orbital migration for $\sim 1$\,Gyr if they had a primordial atmosphere.
The atmospheric escape of super-Earths by photoevaporation can affect the orbital configuration of compact, multi-planet systems, such as the TRAPPIST-1 and Kepler-186 systems.
The orbital separations between each planet in these systems are $\sim 10\,\%$ to their planetary orbital radius, while a planet with an initial atmosphere of 10\,wt\% increases its orbital radius by approximately 10\,\% via atmospheric escape. Planets can be kicked out or pushed into mean motion resonances.

We performed population synthesis studies on the orbital evolution of an evaporating super-Earth around FGKM-type stars.
Close-in planets with $P \lesssim$ 3--5\,days around FGK-type stars lose almost all of the H$_2$/He atmospheres and then move outward, which is consistent with the ``Neptune desert'' in the observed planet population. 
We found that there is a radius gap near 1.6--1.8\,$R_\oplus$, as seen in observations, in the size distribution of planets around FGK-type stars after the orbital migration via atmospheric escape occurs.
This supports the hypothesis that the radius gap reflects the bimodal distribution of the amount of atmosphere, which is caused by atmospheric escape \citep[e.g.,][]{2017ApJ...847...29O}.
The period-radius distribution of close-in super-Earths is highly sensitive to a stellar XUV luminosity, the composition of a rocky core, the initial atmospheric mass fraction, and the range of the orbital period.
The observed planet population around M-type stars can be reproduced only by a high stellar XUV luminosity model.
The detection of more super-Earths around M-type stars by high-precision near-infrared Doppler surveys and the TESS mission will enable us to verify our photoevaporation model.

\acknowledgments

We gratefully acknowledge the referee's helpful comments.
We thank Teruyuki Hirano for his valuable advice on the selection of observational data.
We also thank Ryan Cloutier for providing us with useful data on planet occurrence rates around low-mass stars.
This work was supported by JSPS KAKENHI Grant Number 19K03950. YH was supported by a Grant-in-Aid for Scientific Research on Innovative Areas (JSPS KAKENHI Grant Number 18H05439).

\appendix

\section{TRANSITION OF ATMOSPHERIC ESCAPE REGIME}
\label{appendix:transition}

\begin{figure}[ht!]
\centering
\includegraphics[width=\linewidth,clip]{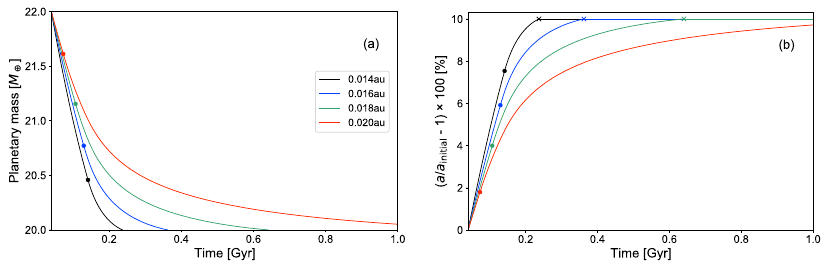}
\caption{Time evolution of (a) mass and (b) orbital radius of a 20\,$M_\oplus$ super-Earth.
The filled circles represent the transition point from radiation recombination-limited to the energy-limited regime.
The planets initially have a H$_2$/He atmosphere of 10\,\% of their core mass at different initial locations: 0.014 (black lines), 0.016 (blue lines), 0.018 (green lines), and 0.020\,au (red lines) around a G-type star.
Crosses represent the point that a H$_2$/He atmosphere is completely lost.
$\chi$ is assumed to be 0.5.
Planets closer to their host stars undergo the transition from the radiation recombination-limited to the energy-limited regime in the later stage because of a higher XUV flux.}
\label{fig:recombination}
\end{figure}

Two mechanisms of atmospheric escape from a planet depend on the core mass and H$_2$/He atmospheric mass. The pressure, density, and velocity of the X-ray heated flow change steeply in the subsonic region from the photoionization base (i.e., the planetary radius) to the sonic surface. The properties of the sonic point for the evaporative wind in the radiation recombination-limited regime are sensitive to the planetary mass and radius. The transition from the radiation recombination-limited to the energy-limited regime occurs when the photoionization base, which corresponds to $R_\mathrm{XUV}$, is equal to the sonic point. If a planetary core mass is sufficiently small ($\lesssim 20\,M_\oplus$), $r_\mathrm{s}$ can become smaller than the core radius $R_\mathrm{core}$. We adopt the energy-limited regime if $r_\mathrm{s} < R_\mathrm{core}$ because the radiation recombination-limited regime is not expected to occur.

Figure \ref{fig:recombination} demonstrates that a 20\,$M_\oplus$ planet undergoes the transition from the radiation recombination-limited to the energy-limited regime for mass loss. For planets closer to their host stars, the transition from the radiation recombination-limited to the energy-limited regime occurs in the later stage because of a higher flux of stellar XUV photons.
They lose a large fraction of the H$_2$/He atmosphere in the radiation recombination-limited regime.
The radiation recombination processes should be effective in the upper atmosphere of close-in planets, especially, close-in massive planets.

\section{Tidal Evolution of a Super-Earth} \label{appendix:tide}

\begin{figure}[ht!]
\centering
\includegraphics[width=\linewidth,clip]{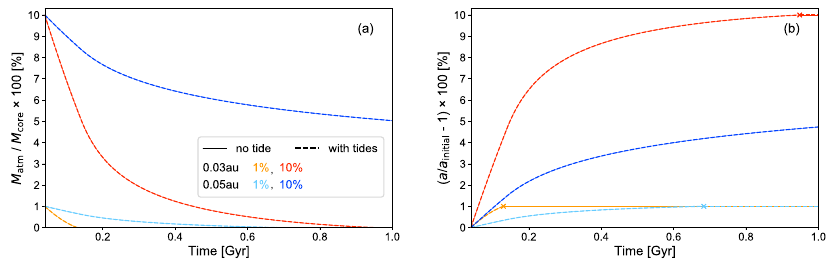}
\caption{Time evolution of the atmospheric mass fraction and the orbital radius of super-Earths without and with tidal effects (thin solid and thick dashed lines). A 10\,$M_\oplus$ super-Earth around a G-type star initially has a 1\,wt\% and 10\,wt\% H$_2$/He atmosphere at 0.03\,au (orange and red) and 0.05\,au (cyan and blue). 
Crosses represent the point that a H$_2$/He atmosphere is completely lost.
$\chi$ is assumed to be 0.5.}
\label{fig:tide}
\end{figure}

In a star-planet system, the orbital change of a planet is mainly caused by the tidal deformation of its host star and the tidal dissipation itself. The orbital decay rate of a planet by tidal interactions \citep{2009ApJ...705L..81V} can be written as
\begin{equation}
  \left(\frac{\dot{a}}{a}\right)_\mathrm{tide,\star} = \frac{f}{\tau_d} \frac{M_\mathrm{env}}{M_\star} q(1+q) \left(\frac{R_\star}{a}\right)^8,
  \label{eq:tide_adot}
\end{equation}
where $q = M_\mathrm{p}/M_\star$, $M_\mathrm{p}$ is the mass of a planet, $a$ is the semimajor axis of the planet, $M_\star$ and $R_\star$ are the mass and radius of a star, $L_\star$ is the stellar luminosity, $M_\mathrm{env}$ is the mass enclosed in the convective envelope of a star, $\tau_d$ is the eddy turnover timescale \citep{1996ApJ...470.1187R} given by
\begin{equation}
  \tau_d = \left[\frac{M_\mathrm{env} (R_\star-R_\mathrm{env})^2}{3 L_\star} \right]^{1/3}.
  \label{eq:tide_tau_d}
\end{equation}
The dissipation factor $f = (P/2 \tau_d)^2$ when $\tau_d > P/2$; otherwise, we take $f = 1$, where $P$ is the orbital period of the planet.
Equation (\ref{eq:tide_adot}) gives us 
\begin{equation}
    \tau_{a,\star} \sim 3 \times 10^{13} \left(\frac{f}{0.01}\right)^{-1}  \left(\frac{\tau_d}{10\,\mathrm{days}}\right) \left(\frac{M_\mathrm{env}/M_\star}{0.025}\right)^{-1} \left(\frac{q}{10^{-5}}\right)^{-1}
  \left(\frac{R_\star}{R_\odot}\right)^{-8} \left(\frac{a}{0.03\,\mathrm{au}}\right)^8
  \,\,\mathrm{yr},
  \label{eq:tau_a1}
\end{equation}
where $\tau_{a,\star} \equiv (a/\dot{a})_\mathrm{tide,\star}$.
This indicates that a close-in super-Earth around a Sun-like star can be locked into a synchronous rotation by tidal forces in $> 1$\,Gyr.
The orbit of a super-Earth around an M-type star decays more slowly because a low-mass star has a fully convective interior and a small radius.

Tidal dissipation in a planet also damps orbit and eccentricity. The orbital decay rate is given by 
\begin{equation}
    \left(\frac{\dot{a}}{a}\right)_\mathrm{tide,p} 
       = \frac{2e^2}{1-e^2} \left(\frac{\dot{e}}{e}\right)_\mathrm{tide,p} = 
           \frac{2a}{GM_\star M_\mathrm{p}} \dot{E},
    \label{eq:tide-dissipation}
\end{equation}

where $e$ is the eccentricity and $\dot{E}$ is the energy loss due to tidal dissipation given by $\dot{E} \sim -(63e^2\Omega_\mathrm{p}/4Q_\mathrm{p} \tilde{\mu}_\mathrm{p})(R_\mathrm{p}/a)^5 GM^2_\star/a$ when $e \ll 1$ \citep[e.g.][]{1966Icar....5..375G}, where $R_\mathrm{p}$ is the planetary radius, $Q_\mathrm{p}$ is the tidal dissipation factor of the planet, and $\tilde{\mu}_\mathrm{p}$ is the bulk modulus of the planet. Note that the tidal circularization of a planet occurs prior to orbital decay when $0 < e \ll 1$. From above the equations, we yield
\begin{equation}
    \tau_{a,\mathrm{p}}\sim 2 \times 10^{11} \left(\frac{Q_\mathrm{p}}{100}\right)
              \left(\frac{k_{2,\mathrm{p}}}{0.3}\right)
              \left(\frac{q}{10^{-5}}\right)
              \left(\frac{M_\star}{M_\odot}\right)^{-\frac{1}{2}}
              \left(\frac{R_\mathrm{p}}{R_\oplus}\right)^{-5}
              \left(\frac{a}{0.03\,\mathrm{au}}\right)^{\frac{13}{2}}
              \left(\frac{e}{0.01}\right)^{-2}\,\,\mathrm{yr},
    \label{eq:tau_a2}
\end{equation}
where $\tau_{a,\mathrm{p}} \equiv (a/\dot{a})_\mathrm{tide,\mathrm{p}}$, and Love number $k_{2,\mathrm{p}} = \frac{3}{2}\frac{1}{1 + \tilde{\mu_p}}$.
The tidal decay of a close-in super-Earth around FGKM-type stars due to tidal dissipation has a negligible effect on the orbital evolution for $\lesssim 1$\,Gyr.

Figure \ref{fig:tide} demonstrates the orbital evolution of a 10\,$M_\oplus$ super-Earth around a G-type star due to mass loss and tidal interactions.
The size and heat flux in the stellar convective envelope vary with time.
The turnover timescale in the convective zone was calculated using the stellar evolution code {\tt MESA} \citep{2011ApJS..192....3P}.
A planet initially has a 1\,wt\% and a 10\,wt\% H$_2$/He atmosphere at 0.03\,au and 0.05\,au in a circular orbit. The orbital evolution of a planet in the absence of tidal interactions is also shown in Figure \ref{fig:tide}. We can see that tidal forces hardly change the orbital evolution of a planet for 1\,Gyr. The orbital decay of a planet due to tidal forces had little effect on our simulations.

\bibliography{main}{}
\bibliographystyle{aasjournal}

\end{document}